%% file: main.tex
\documentclass[11pt]{article}
\usepackage[margin=1in]{geometry}
\usepackage[T1]{fontenc}
\usepackage{lmodern}
\usepackage{microtype}
\usepackage{amsmath,amssymb}
\usepackage{booktabs}
\usepackage{graphicx}
\usepackage{array}
\usepackage{tabularx}
\usepackage[table]{xcolor}
\usepackage{listings}
\usepackage{caption}
\usepackage[section]{placeins}

\usepackage[hidelinks]{hyperref}
\usepackage{cleveref}

\input{macros}
\title{Do Apple M-Series Processors Enforce the  Compound Memory Model?\\[0.5ex]\large Sequential and Parallel Composition of Arm and TSO}
\author{Vijay Nagarajan\\University of Utah\\\texttt{vijay@cs.utah.edu}}
\date{}

\begin{document}
\maketitle

\begin{abstract}
\input{sections/abstract}
\end{abstract}

\input{sections/intro}
\input{sections/models}
\input{sections/compose}

\input{sections/method}
\input{sections/results-arm}
\input{sections/results-tso}
\input{sections/results-compound}
\input{sections/results-scale}
\input{sections/conclusion}
\input{sections/artifacts}

\section*{Acknowledgements}
The M1 runs used cfarm103 of the GCC Compile Farm; we thank its maintainers. The per-thread TSO interface used here is the work of the Asahi Linux developers. This work is supported in part by the National Science Foundation under grant CCF-2525270, and by Arm.
This work was carried out with substantial assistance from Claude (Anthropic) in preparing and running the tests, writing the herd models, developing the tables and figures, and writing the report. The author takes full responsibility for the contents of this report.

\bibliographystyle{plain}
\bibliography{refs}
\end{document}

%% file: macros.tex
\newcommand{\rel}[1]{\ensuremath{\mathsf{#1}}}
\newcommand{\po}{\rel{po}}
\newcommand{\poloc}{\rel{po\text{-}loc}}
\newcommand{\rf}{\rel{rf}}
\newcommand{\rfe}{\rel{rfe}}
\newcommand{\co}{\rel{co}}
\newcommand{\coe}{\rel{coe}}
\newcommand{\fr}{\rel{fr}}
\newcommand{\fre}{\rel{fre}}
\newcommand{\rmw}{\rel{rmw}}
\newcommand{\ppo}{\rel{ppo}}
\newcommand{\armppo}{\rel{ppo}_{\mathrm{Arm}}}
\newcommand{\tsoppo}{\rel{ppo}_{\mathrm{TSO}}}
\newcommand{\ins}[1]{\texttt{#1}}

\newcommand{\rd}[1]{\quad\textit{(reads #1)}}

\newcommand{\newtext}[1]{#1}
\newcommand{\rev}[1]{#1}  %
\newcommand{\newer}[1]{#1}  %
\newcommand{\audit}[1]{#1}  %
\newcommand{\rvb}[1]{#1}  %
\newcommand{\rvc}[1]{#1}  %
\newcommand{\rvd}[1]{#1}  %
\newcommand{\rve}[1]{#1}  %

\lstdefinestyle{cat}{basicstyle=\ttfamily\footnotesize,columns=fullflexible,keepspaces=true,
  frame=lines,morecomment=[s]{(*}{*)},commentstyle=\itshape\color{gray}}



\definecolor{armc}{HTML}{0072B2}
\definecolor{tsoc}{HTML}{D55E00}
\colorlet{armbg}{armc!9}
\colorlet{tsobg}{tsoc!11}
\newcommand{\armw}[1]{\textcolor{armc}{#1}}
\newcommand{\tsow}[1]{\textcolor{tsoc}{#1}}

\newcommand{\thr}[2]{Thread #2 (\textcolor{#1c}{\textbf{\csname#1name\endcsname}})}
\newcolumntype{A}{>{\columncolor{armbg}[0pt][\tabcolsep]}l}
\newcolumntype{T}{>{\columncolor{tsobg}[0pt][\tabcolsep]}l}
\newcolumntype{B}{>{\columncolor{armbg}[\tabcolsep][0pt]}l}
\newcolumntype{U}{>{\columncolor{tsobg}[\tabcolsep][0pt]}l}
\newcommand{\litmusrules}{\setlength{\aboverulesep}{0pt}\setlength{\belowrulesep}{0pt}\renewcommand{\arraystretch}{1.15}}

\definecolor{okc}{HTML}{009E73}
\newcommand{\okbox}[1]{{\setlength{\fboxsep}{1.2pt}\colorbox{okc!16}{#1}}}
\newcommand{\exbox}[1]{{\setlength{\fboxsep}{1.2pt}\colorbox{black!24}{#1}}}
\newcommand{\rfz}{\okbox{0}}                               %
\newcommand{\rsn}[1]{\okbox{#1}}                           %
\newcommand{\rst}[1]{\exbox{\textbf{0}\,(#1)}}             %
\newcommand{\rux}{\exbox{\textbf{0}$^{\dagger}$}}          %

%% file: sections/abstract.tex
Apple's M-series processors can run each thread under the Arm memory model or under a TSO model, and threads of both kinds can share memory. This raises two questions of composition that Apple has not answered. Across threads, how does the machine behave as a whole? Within a thread, which Arm instructions play the roles of TSO's loads, stores and atomic read-modify-writes, and how are they ordered between themselves and across other Arm instructions?

We propose that both answers are compositions of preserved program order. Across threads, each thread keeps the preserved program order of its own model: a parallel composition. Within a TSO thread, every Arm instruction keeps TSO's order and adds Arm's, and every atomic memory operation (AMO) is locked, as x86's read-modify-writes are: a sequential composition. 

We test both on an Apple M1 and an Apple M2 running Linux, with 936 litmus tests chosen from the two models' preserved program order, each run $6\times10^8$ times on each machine.
Arm-mode threads follow Arm's official model, TSO-mode threads follow x86-TSO and the sequential composition, and programs that mix the two modes follow the compound model: neither machine shows an outcome these models forbid. Both machines also preserve three orders that these models do not require, and a compare-and-swap whose comparison fails behaves as if it writes back the value it read, which the Arm model does not require. All three orders held in runs up to 40 times longer.

%% file: sections/intro.tex
\section{Introduction}
\label{sec:intro}

Apple's M-series processors can run any thread under the Arm memory model or under a TSO memory model, and Arm threads and TSO threads can run side by side. This raises two questions of composition. Across threads, the two kinds of thread share memory, and the machine as a whole needs a memory model: a \emph{parallel composition}. Within a thread, a TSO thread on an Apple core runs Arm instructions that TSO does not define, such as acquire loads, release stores and Arm's barriers, and a rule must say how they are ordered with the thread's other accesses: a \emph{sequential composition}. Apple has specified neither\newer{: its documentation says only that the mode bit changes a core's memory model to ``the x86-64 TSO memory model''~\cite{apple-rosetta-tso}, which defines x86's own operations but not the Arm instructions x86 lacks, nor threads of both kinds sharing memory}. 

There is demand for this specification. 
The Asahi Linux kernel exposes TSO mode per thread, and the x86 emulator FEX-Emu uses it~\cite{lkml-tso,fex-2306}\newer{; Apple distributes a Linux patch with the same interface for its translator, Rosetta, in Linux virtual machines~\cite{apple-rosetta-tso}}; the same interface was declined for mainline Linux, in part because no specification says which memory model the mode implements~\cite{lkml-tso-nak,lkml-tso-spec}. \newer{Wrenger et al.\ measured the performance cost of TSO mode on an M1 Ultra~\cite{wrenger-jsa24}, and Khayet and Morrison found it typically within 3\% and traced the larger slowdowns to Apple's implementation of the mode~\cite{khayet-asplos26}; this report concerns which orders the mode preserves.} This report proposes a model for each composition and tests both on an Apple M1 and an Apple M2.

\begin{figure}[!htbp]
\centering
\footnotesize
\litmusrules
\newcommand{\pt}[1]{\multicolumn{2}{@{}l}{\textbf{#1}}\\[2pt]}
\begin{minipage}[t]{0.48\linewidth}\centering
\begin{tabular}{@{}AB@{}}
\pt{(a) MP+rel+acq}
\toprule
\thr{arm}{0} & \thr{arm}{1}\\
\midrule
\ins{x = 1} & \ins{r1 = acq ld y}\rd{1}\\
\ins{rel y = 1} & \ins{r2 = ld x}\rd{0}\\
\bottomrule
\end{tabular}
\end{minipage}\hfill
\begin{minipage}[t]{0.48\linewidth}\centering
\begin{tabular}{@{}TU@{}}
\pt{(b) MP}
\toprule
\thr{tso}{0} & \thr{tso}{1}\\
\midrule
\ins{x = 1} & \ins{r1 = ld y}\rd{1}\\
\ins{y = 1} & \ins{r2 = ld x}\rd{0}\\
\bottomrule
\end{tabular}
\end{minipage}

\vspace{3ex}
\begin{minipage}[t]{0.48\linewidth}\centering
\begin{tabular}{@{}AU@{}}
\pt{(c) MP+rel}
\toprule
\thr{arm}{0} & \thr{tso}{1}\\
\midrule
\ins{x = 1} & \ins{r1 = ld y}\rd{1}\\
\ins{rel y = 1} & \ins{r2 = ld x}\rd{0}\\
\bottomrule
\end{tabular}
\end{minipage}

\vspace{3ex}
\begin{minipage}[t]{0.48\linewidth}\centering
\begin{tabular}{@{}TU@{}}
\pt{(d) SB+rel+acq}
\toprule
\thr{tso}{0} & \thr{tso}{1}\\
\midrule
\ins{rel x = 1} & \ins{rel y = 1}\\
\ins{r1 = acq ld y}\rd{0} & \ins{r2 = acq ld x}\rd{0}\\
\bottomrule
\end{tabular}
\end{minipage}\hfill
\begin{minipage}[t]{0.48\linewidth}\centering
\begin{tabular}{@{}TU@{}}
\pt{(e) SB+acq}
\toprule
\thr{tso}{0} & \thr{tso}{1}\\
\midrule
\ins{x = 1} & \ins{y = 1}\\
\ins{r1 = acq ld y}\rd{0} & \ins{r2 = acq ld x}\rd{0}\\
\bottomrule
\end{tabular}
\end{minipage}
\caption{Initially \ins{x = y = 0}; \ins{rel} is \ins{STLR}, \ins{acq ld} is \ins{LDAR}. Columns of Arm threads are shaded blue, of TSO threads orange. Top two rows: message passing, the parallel composition. The outcome shown is forbidden by Arm in (a), by TSO in (b), and by the compound model in (c), where Thread~0 keeps Arm's order of its two stores and Thread~1 TSO's order of its two loads. Bottom row: store buffering on TSO threads, the sequential composition. x86-TSO allows the outcome of both, since it does not order a store before a later load. The sequential composition forbids (d), because Arm orders a release store before a later \ins{LDAR}, and allows (e). Neither machine shows (a)--(d), and both show (e), $10^6$ times on the M1 and $10^5$ on the M2, in $6\times10^8$ runs of the tests of \cref{sec:arm,sec:tso,sec:compound}; (b) and (c) are run as shown, and in (a), (d) and (e) one thread's annotations are replaced by a \ins{DMB}.}
\label{fig:intro}
\end{figure}

\paragraph{Parallel composition.} A number of recent works propose the compound memory model for machines on which two memory models share memory~\cite{heterogen,cmcm-pldi23}. A compound model is not a new memory model: each thread preserves the program order that its own memory model preserves, and the rest of the model is shared by all threads. If the machine enforces it, each ISA's compiler mappings, and hence each language's implementation of its synchronization primitives, stay correct after the composition~\cite{cmcm-pldi23}. The NVIDIA Grace-Hopper, which places Arm CPU cores and a PTX GPU on one coherent memory, adheres to the compound model of Arm and PTX for message-passing litmus tests~\cite{gracehopper-ismm26}. Apple's processors are a more direct test, because the same core runs a thread as Arm or as TSO. \Cref{fig:intro}(a) and (b) show message passing under each model alone; in (c) the producer is an Arm thread and the consumer a TSO thread. The compound model forbids the outcome by taking each thread's order from its own model: the release orders the producer's stores, and TSO orders the consumer's loads. With a plain store in place of the release, or with the consumer in Arm mode, the outcome is allowed.

\paragraph{Sequential composition.} A TSO thread on an Apple core runs Arm instructions. An Arm load or store, whatever its annotation, is \rvb{at least} a TSO load or store, and the full \ins{DMB} is \ins{MFENCE}. Acquire loads and release stores also carry orders x86 does not have, and the barriers \ins{DMB LD}, \ins{DMB ST} and \ins{DSB} have no x86 counterpart. We propose that an Arm instruction on a TSO thread keeps TSO's order and adds Arm's: two accesses are ordered if either model orders them. TSO orders every pair of accesses except a store followed by a load, so Arm's orders can add only to such a pair. On a TSO thread, a release store is therefore also ordered before a later \ins{LDAR}, which a plain store is not. \Cref{fig:intro}(d) and (e) show the difference: with acquire loads, store buffering is forbidden with release stores and allowed with plain ones. We also propose that every atomic memory operation (AMO) is locked, as an x86 read-modify-write (RMW) is, and so acts as a fence. Arm's other RMW, a load-exclusive followed by a store-exclusive (an LL/SC pair), is atomic but not locked (\cref{sec:models-seq}). We propose that the compound model of Arm and TSO is a parallel composition, whose TSO side is itself a sequential composition.

\paragraph{Questions.} Whether Apple's processors follow this model comes down to three questions: one for each mode, and one for programs that mix them.
\begin{enumerate}
\item Does Arm mode follow Arm's official model~\cite{arm-arm,arm-pulte}?
\item Which TSO does TSO mode implement, and does it follow the sequential composition? First, on the operations x86 has: is it x86-TSO~\cite{x86tso-cacm}, in which a load may read its own thread's store from the store buffer before other threads see it, or a stronger variant such as the IBM 370 model~\cite{adve-tutorial}, in which it may not? Second, on a TSO-mode thread, do the other Arm instructions behave as the sequential composition predicts?
\item Do programs that mix Arm-mode and TSO-mode threads follow the compound model?
\end{enumerate}

\paragraph{Method.} These questions were first examined around 2024, on an M1 Mac mini under macOS, using the third-party TSOEnabler kernel extension to put threads into TSO mode~\cite{pldi26}\rvb{, which reported the outcome forbidden by the compound model in mixed message passing in both directions: an Arm producer with \ins{DMB ST} and a TSO consumer, and a TSO producer and an Arm consumer with \ins{DMB LD}}. Here we use the Asahi Linux interface instead, on an M1 Mac mini and an M2 MacBook Air, and run 936 litmus tests with an otherwise unmodified litmus7~\cite{litmus7} (\cref{sec:method}).
The tests are chosen from the preserved program order of Arm and of the sequential composition. Every row of each is tested in a two-thread test, with the other thread in each mode. Three- and four-thread tests run under every assignment of modes, with the doubtful mechanisms (\cref{sec:method-tests}): on Arm threads, dependencies, acquire loads, release stores and \ins{DMB ST}; on TSO threads, plain accesses, \ins{DMB} as \ins{MFENCE}, and acquire loads and release stores. Further tests ask whether a thread reads its own store early, and whether RMWs are atomic and which are locked. Every test runs $6\times10^8$ times on each machine.
\rvb{Our suite includes both tests reported in~\cite{pldi26}. Neither machine showed the outcome in $6\times10^8$ runs of either test, although it appeared in both once the barrier was removed (\cref{sec:compound}).} The two setups differ in operating system and in how the TSO bit is set: macOS with a kernel extension there, Linux with a per-thread system call here.

\paragraph{Answers.}
\begin{enumerate}
\item Yes. No outcome the Arm model forbids was observed in Arm mode on either machine.
\item TSO mode implements x86-TSO rather than a stronger variant: a load reads its own thread's store early, as x86-TSO allows and IBM 370 does not. On a TSO-mode thread the other Arm instructions follow the sequential composition.
\item Yes. No outcome the compound model forbids was observed on either machine.
\end{enumerate}
Both machines are stronger than these models in two ways. They preserve three orders that none of the models requires: an acquire load never reads its own thread's store early, \ins{DMB LD} and \ins{DSB LD} order a store before every later access, and on Arm threads a load is ordered before a later store. \rvb{A fourth order held in all $6\times10^8$ runs: the write of an acquire AMO is ordered before later accesses (\cref{fig:rmw-types}(b)). In runs about 40 times longer, however, the order was violated on both machines, albeit rarely (\cref{sec:scale}), so we do not add it.} And a \ins{CAS} whose comparison fails behaves as if it writes back the value it read, which makes a failing \ins{CASAL} a fence (\cref{sec:arm}). These describe the two machines, not what either ISA promises, so we add them to separate models of the machines (\cref{sec:models-apple}) and leave the proposal unchanged.

%% file: sections/models.tex
\section{Background: Arm and x86-TSO}
\label{sec:models}
\label{sec:background}

This section recalls the Arm and x86-TSO memory models. \Cref{sec:compose} proposes a theory of their sequential and parallel composition, and \cref{sec:arm,sec:tso,sec:compound} test it.

\subsection{Executions}

We use the axiomatic framework of Alglave et al.~\cite{herding-cats}. An execution is a set of memory events (reads R, writes W and fences) with the usual relations: program order \po, reads-from \rf, coherence order \co, and from-read $\fr = \rf^{-1};\co$. A suffix \textsf{e} restricts a relation to events of different threads, as in \rfe, \coe\ and \fre. The relation \rmw\ pairs the read and the write of a read-modify-write (RMW, \cref{sec:rmw}).

\subsection{Preserved Program Order}

Informally, Arm and x86-TSO both present the threads with an atomic memory: once a write is visible to another thread, it is visible to every thread. (Formally, both are other-multi-copy atomic, \cref{sec:axioms}.) What distinguishes them is which pairs of a thread's accesses they keep in program order, their \emph{preserved program order} (\ppo). A pair not in \ppo\ may appear to other threads in either order. We include in \ppo\ every \emph{mechanism} a model provides for ordering two accesses of a thread: annotations on the accesses, barriers and fences between them, dependencies, and RMWs; and, in TSO, the orders it keeps by default. \Cref{tab:ppo} lists them for accesses to different addresses. Every model in this report, the compositions included, has the same axioms and differs from the others only in \ppo.

\begin{table}[!htbp]
\centering
\small
\begin{tabularx}{\linewidth}{@{}l>{\raggedright\arraybackslash}X>{\raggedright\arraybackslash}p{0.25\linewidth}@{}}
\toprule
Order & \armw{Arm} & \tsow{x86-TSO}\\
\midrule
load $\to$ load & first is acquire; \ins{DMB}, \ins{DMB LD}, \ins{DSB} or \ins{DSB LD} between; second address-dependent on first, or after a control dependency and \ins{ISB} & always\\
load $\to$ store & first is acquire; second is release; \ins{DMB}, \ins{DMB LD}, \ins{DSB} or \ins{DSB LD} between; second dependent on first & always\\
store $\to$ store & second is release; \ins{DMB}, \ins{DMB ST}, \ins{DSB} or \ins{DSB ST} between & always\\
store $\to$ load & first is release and second is \ins{LDAR} or \ins{LDAXR}; \ins{DMB}, \ins{DSB} or \ins{DSB ST} between & \ins{MFENCE} or a locked RMW between; either is part of a locked RMW\\
\bottomrule
\end{tabularx}
\caption{Preserved program order (\ppo) in Arm and x86-TSO, for a first access and a later second access to a different address. 
RMWs add the orders of \cref{sec:rmw}.}
\label{tab:ppo}
\end{table}

\paragraph{Arm.} A release store is ordered after every earlier access, and an acquire load before every later access; a release store is also ordered before a later \ins{LDAR} (an RCsc acquire), though not before a later \ins{LDAPR} (an RCpc acquire)~\cite{arm-pulte}. The full barriers \ins{DMB} and \ins{DSB} order every earlier access before every later one. \ins{DMB LD} and \ins{DSB LD} order earlier loads before every later access; \ins{DMB ST} orders earlier stores before later stores, and \ins{DSB ST} orders earlier stores before every later access. Address, data and control dependencies order a load before the dependent access, in the cases the architecture lists. Any other pair of accesses to different addresses may be reordered.

\paragraph{x86-TSO.} Every pair is preserved except a store followed by a load~\cite{x86tso-cacm}. An \ins{MFENCE} between the two restores that order, and so does a locked RMW, between them or as one of the two (\cref{sec:rmw}).

\subsection{The Axioms}
\label{sec:axioms}

The three axioms are the following; the third depends on \ppo,\  which differs for x86-TSO and Arm. 
\begin{align*}
&\textsf{acyclic}(\poloc \cup \rf \cup \co \cup \fr) && \text{(sc-per-location)}\\
&\rmw \cap (\fre;\coe) = \emptyset && \text{(atomicity)}\\
&\textsf{acyclic}(\ppo \cup \rfe \cup \coe \cup \fre) && \text{(order)}
\end{align*}
The first says that the accesses to each address agree with a total order on that address. The second says that no other thread's write comes between the read and the write of an RMW. The third says that the orders a thread preserves are the orders other threads observe: \ppo, together with communication between threads, never forms a cycle. herd's x86-TSO model writes \co\ and \fr\ in full in the third axiom; the internal parts are already in TSO's \ppo, so the two forms agree.

The order axiom uses \rfe\ rather than \rf, and this fixes the models' degree of multi-copy atomicity. There are three degrees. In a \emph{multi-copy atomic} model, such as the IBM 370~\cite{adve-tutorial}, a write becomes visible to all threads at the same time, its own thread included; the order axiom would use \rf. In an \emph{other-multi-copy atomic} model, a thread may read its own write early, from its store buffer, but once the write is visible to one other thread it is visible to all of them; the order axiom uses \rfe. Arm and x86-TSO are both of this kind. In a non-multi-copy atomic model, such as Power, other threads may see a write at different times. We do not consider such models.

\Cref{fig:models-tests} illustrates the two halves of other-multi-copy atomicity. IRIW asks whether two readers can see two independent writes in opposite orders. Neither model allows it once each reader's two loads are ordered: TSO orders them by default, Arm when they are acquire loads. Arm allows the outcome with plain loads only because it may reorder them. n6~\cite{x86tso-cacm} asks whether a thread may read its own write before other threads see it. Thread~0 reads its own write of \ins{x} and then reads \ins{y} as 0, while Thread~1's write \ins{x = 2} precedes \ins{x = 1} in coherence. Both models allow this. \rvb{In TSO,} Thread~0 may read \ins{x = 1} from its store buffer before the write is visible to Thread~1\rvb{, and since TSO orders every other pair of accesses here, this is the only way to get the outcome. (Arm allows it even without this, since it may reorder Thread~0's loads.)} With \rf\ in place of \rfe\ in the order axiom, \rvb{TSO} forbids the outcome.

\begin{figure}[!htbp]
\centering
\small
\begin{tabular}{@{}llll@{}}
\multicolumn{4}{@{}l}{\textbf{(a) IRIW}}\\
\toprule
Thread 0 & Thread 1 & Thread 2 & Thread 3\\
\midrule
\ins{x = 1} & \ins{r1 = ld x}\rd{1} & \ins{y = 1} & \ins{r3 = ld y}\rd{1}\\
 & \ins{r2 = ld y}\rd{0} & & \ins{r4 = ld x}\rd{0}\\
\bottomrule
\end{tabular}

\vspace{1ex}
\begin{tabular}{@{}lll@{}}
\toprule
Threads & Readers & Verdict\\
\midrule
all \armw{Arm} & plain loads & allowed\\
all \armw{Arm} & acquire loads & forbidden\\
all \tsow{TSO} & plain loads & forbidden\\
\bottomrule
\end{tabular}

\vspace{2ex}
\begin{tabular}{@{}ll@{}}
\multicolumn{2}{@{}l}{\textbf{(b) n6}}\\
\toprule
Thread 0 & Thread 1\\
\midrule
\ins{x = 1} & \ins{y = 2}\\
\ins{r1 = ld x}\rd{1} & \ins{x = 2}\\
\ins{r2 = ld y}\rd{0} & \\
\bottomrule
\end{tabular}
\quad
\begin{tabular}{@{}ll@{}}
\toprule
Model & Verdict\\
\midrule
\armw{Arm} & allowed\\
\tsow{TSO} & allowed\\
\tsow{TSO}, multi-copy atomic & forbidden\\
\bottomrule
\end{tabular}
\caption{Other-multi-copy atomicity. (a) Once each reader's loads are ordered, every thread sees the two writes in one order. (b) A thread may read its own write before other threads see it, which full multi-copy atomicity forbids. Initially all locations are 0; in (b) the final value of \ins{x} is 1. }
\label{fig:models-tests}
\end{figure}

\subsection{Read-Modify-Writes}
\label{sec:rmw}

An RMW is an instruction, or a pair of instructions, that reads a location and writes it atomically. It contributes a read and a write related by \rmw.

\begin{figure}[!htbp]
\centering
\footnotesize
\litmusrules
\begin{minipage}[t]{0.53\linewidth}\centering
\begin{tabular}{@{}AB@{}}
\multicolumn{2}{@{}l}{\textbf{(a) Release RMWs}}\\
\toprule
\thr{arm}{0} & \thr{arm}{1}\\
\midrule
\ins{x = 1} & \ins{y = 1}\\
\ins{r1 = swpl y, 2}\rd{0} & \ins{r2 = swpl x, 2}\rd{0}\\
\bottomrule
\end{tabular}
\end{minipage}\hfill
\begin{minipage}[t]{0.45\linewidth}\centering
\begin{tabular}{@{}AB@{}}
\multicolumn{2}{@{}l}{\textbf{(b) Acquire RMWs}}\\
\toprule
\thr{arm}{0} & \thr{arm}{1}\\
\midrule
\ins{r1 = swpa x, 1} & \ins{r3 = swpa y, 1}\\
\ins{r2 = ld y}\rd{0} & \ins{r4 = ld x}\rd{0}\\
\bottomrule
\end{tabular}
\end{minipage}
\caption{Orders that atomicity gives an RMW. (a) Release RMWs in place of the loads: Arm's type-3 atomicity forbids the outcome, with the release orders, although no rule of preserved program order orders a store before a later RMW's read. (b) Acquire RMWs in place of the stores: Arm allows the outcome; a type-1 RMW, such as x86's locked RMW, forbids it. Initially all locations are 0. }
\label{fig:rmw-types}
\end{figure}

\paragraph{x86.} The RMWs are the locked instructions: \ins{XCHG} with a memory operand, and the read-modify-write instructions that take a \ins{LOCK} prefix (\ins{LOCK CMPXCHG}, \ins{LOCK XADD}, \ins{LOCK ADD}, etc.). The prefix is allowed only on instructions that both read and write memory, so there is no locked load or locked store, and every locked instruction writes: a \ins{CMPXCHG} whose comparison fails writes back the value it read~\cite{intel-sdm}. In the terms of~\cite{fastrmw-pldi13}, a locked RMW is type-1: no memory event comes between its read and its write. With TSO's \ppo\ this makes it a full fence.

\paragraph{Arm.} There are two kinds, each with optional acquire, release, or both. An \emph{AMO} (atomic memory operation) is a single instruction: \ins{SWP}, \ins{CAS}, or one of the \ins{LD}\textit{op} instructions (\ins{LDADD}, \ins{LDSET}, \ins{LDCLR}, etc.). An \emph{LL/SC pair} is a load-exclusive followed by a store-exclusive (\ins{LDXR} and \ins{STXR}, with \ins{LDAXR} and \ins{STLXR} as the acquire and release forms); the store-exclusive fails if another write to the location may have intervened, and the program then usually retries the pair. \ins{SWP} and \ins{LD}\textit{op} always write. A \ins{CAS} whose comparison fails either writes back the value it read or does not write at all; Arm leaves the choice to the implementation~\cite{arm-a64-isa}.

Arm gives both kinds the same atomicity axiom, above, which in the terms of~\cite{fastrmw-pldi13} is type-3: other threads' writes to the location cannot come between the read and the write, but their reads can. Arm adds three orders for RMWs. The read of an RMW is ordered before its write. If a later acquire load of the same thread reads the RMW's own write, the RMW's read is ordered before that load. And the write of an AMO with both acquire and release (\ins{SWPAL}, \ins{LDADDAL}, etc.) is ordered before every later access, which with the release order makes such an AMO a full fence.

Atomicity also orders accesses that no rule of preserved program order names. Rajaram et al.~\cite{fastrmw-pldi13} first observed such orders, for TSO, and they matter for Arm too, as \cref{fig:rmw-types} shows. In (a), each thread stores and then runs a release RMW that reads 0. The release orders the store before the RMW's write. Because the RMW read 0, atomicity puts the other thread's store to that location after the RMW's write in coherence. The two release orders and the two coherence edges form a cycle, so the outcome is forbidden, although no rule orders the store before the RMW's read. The same holds whenever something orders a store before an RMW's write, such as a \ins{DMB ST} between them or, on a TSO thread, TSO's order of two stores; for an LL/SC pair, only if the store-exclusive succeeds. In (b), each thread runs an acquire RMW and then a load. The acquire orders the RMW's read before the load but not its write, and type-3 atomicity does not help, since it keeps other threads' writes, not their reads, out from between the RMW's read and write. Arm therefore allows the outcome. Recall that in a type-1 RMW, such as x86's locked RMWs, nothing comes between the read and the write, so the load, ordered after the read, is also ordered after the write, and the outcome is forbidden.

Throughout the report, \emph{RMW} means either kind, and \emph{AMO} and \emph{LL/SC} distinguish them.

%% file: sections/compose.tex
\section{Composing Arm and TSO}
\label{sec:compose}

On a thread in TSO mode, an Apple core runs plain loads and stores, which presumably have their x86-TSO meaning, but also Arm instructions x86 does not have, such as acquire loads and release stores. How are these ordered with respect to the plain accesses and to each other? And how do the instructions of a thread in Arm mode interact with those of a thread in TSO mode? We answer the first with the sequential composition, which gives the preserved program order of a TSO thread, and the second with the parallel composition, the compound model, which gives the behavior of a program whose threads run in both modes. We then describe how the models are encoded in herd. We also encode the Apple models, which add the orders both machines are found to preserve \rvd{(three orders and the failing \ins{CAS}, \cref{sec:models-apple})}. 

\subsection{Sequential Composition: Arm Instructions on a TSO Thread}
\label{sec:models-seq}

x86-TSO is defined for four operations: loads, stores, locked RMWs and \ins{MFENCE}. A TSO thread on an Apple core runs Arm instructions instead. Each of the four has Arm instructions that play its role: Arm's loads and stores are TSO loads and stores, an AMO plays the role of a locked RMW, and the full \ins{DMB} is \ins{MFENCE}. Some of these instructions also carry orders that x86 does not have: acquire loads (\ins{LDAPR}, \ins{LDAR}\rvd{, \ins{LDAXR}}) and release stores (\ins{STLR}\rvd{, \ins{STLXR}}). Others have no x86 role at all: the barriers \ins{DMB LD}, \ins{DMB ST} and \ins{DSB}. Neither TSO nor Apple says how these combine with TSO's orders. For every pair of accesses of a TSO thread, in program order, the sequential composition says whether other threads observe the two in that order. It has three parts: 1) every access keeps TSO's order, 2) an AMO is locked, and 3) Arm's orders are added.

\paragraph{TSO's orders.} First, every access of a TSO thread is a TSO access, whatever its annotation. An instruction that reads memory performs a load and one that writes memory performs a store: \ins{LDR}, \ins{LDAPR}, \ins{LDAR} and the load-exclusives are loads, and \ins{STR}, \ins{STLR} and the store-exclusives are stores. TSO orders every pair of these loads and stores except a store followed by a load.

\paragraph{Locked AMOs.} Second, an AMO performs a load followed by a store to the same location, and is locked, as x86's RMWs are. Because TSO already orders every pair of accesses except a store followed by a load, locking adds two orders: every access before the AMO is ordered before the AMO's load, and the AMO's store is ordered before every access after it. Every access before the AMO is then ordered before every access after it, so the AMO acts as a fence. A \ins{CAS} whose comparison fails is also locked. On x86, a failed \ins{LOCK CMPXCHG} writes back the value it read, so it is an ordinary locked RMW. An Arm \ins{CAS} whose comparison fails need not write (\cref{sec:rmw}), but on a TSO thread it is still a fence: a store before it is ordered before its load because it is locked, and its load is ordered before every later access because TSO orders a load before every later access. 

An LL/SC pair is a load and a store, atomic as Arm defines it, but not locked. 
Atomicity still forbids one more outcome. Recall from \cref{sec:rmw} that if a store is ordered before the write of an RMW, the RMW's read behaves as if it were ordered after the store. TSO orders a store before the store-exclusive of a later pair, so when the store-exclusive succeeds, the pair's load cannot be observed before the earlier store. It is not an order of $\tsoppo$, and it does not make the pair a fence: a load after the pair can still be observed before the earlier store (\cref{sec:tso}).

\paragraph{Arm's orders.} Third, we add every order that Arm gives the same instructions (\cref{tab:ppo,sec:rmw}). A pair of accesses of a TSO thread is ordered if TSO orders it or Arm orders it, and otherwise it is not:
\[
\tsoppo = \rel{ppo}_{\mathrm{x86}} \cup \rel{ppo}_{\mathrm{Arm}},
\]
where $\rel{ppo}_{\mathrm{x86}}$ is TSO's order above: every pair of accesses except a store followed by a load, together with a store before a later AMO and an AMO before a later load\rvd{, and a store before a later load across a full \ins{DMB} (\ins{MFENCE})}. We call this the \emph{sequential composition} of x86-TSO and Arm: the two models are composed within one thread, whereas the compound model (\cref{sec:models-par}) composes them across threads.

TSO already orders every pair except a store followed by a load, so Arm's orders can add only to such a pair. Arm orders a store before a later load in two ways: by a full \ins{DMB}, a \ins{DSB} or a \ins{DSB ST} between them, or when the store is a release (\ins{STLR}, \ins{STLXR}) and the load an RCsc acquire (\ins{LDAR}, \ins{LDAXR}). Of these, only the full \ins{DMB} has an x86 counterpart, \ins{MFENCE}, so x86-TSO already has its order. The sequential composition therefore adds to x86-TSO the order of a store before a later load across a \ins{DSB} or a \ins{DSB ST}, and of a release store before a later \ins{LDAR} or \ins{LDAXR}.

The release case needs care, because Arm's release and acquire each give two orders. A release store is ordered after every earlier access, and also before a later \ins{LDAR} or \ins{LDAXR}. An acquire load is ordered before every later access, and, if it is \ins{LDAR} or \ins{LDAXR}, also after an earlier release store. On a TSO thread every store already has the first order of a release, and every load the first order of an acquire. Neither has the second order unless it is annotated, since TSO never orders a store before a later load. So \ins{STLR} followed by \ins{LDAR} is ordered, but \ins{STR} followed by \ins{LDAR} is not, and neither is \ins{STLR} followed by \ins{LDR} or \ins{LDAPR}. \Cref{tab:seq} lists every case.

\begin{table}[!htbp]
\centering
\small
\begin{tabularx}{\linewidth}{@{}Xll@{}}
\toprule
First access, then second & Ordered & By\\
\midrule
load $\to$ load, load $\to$ store, store $\to$ store, any annotations & always & \tsow{TSO}\\
store $\to$ load, either an AMO or an AMO between & yes & \tsow{TSO}: AMOs are locked\\
store $\to$ load, \ins{STLR} or \ins{STLXR} then \ins{LDAR} or \ins{LDAXR} & yes & \armw{Arm}: release, RCsc acquire\\
store $\to$ load, \ins{DMB}, \ins{DSB} or \ins{DSB ST} between & yes & \armw{Arm}: barrier\\
store $\to$ load, otherwise (e.g.\ \ins{STR} then \ins{LDAR}; \ins{STLR} then \ins{LDR} or \ins{LDAPR}; \ins{DMB LD} or \ins{DMB ST} between) & no & \\
\bottomrule
\end{tabularx}
\caption{The sequential composition: preserved program order on a TSO thread, for a first access and a later second access to a different address. }
\label{tab:seq}
\end{table}

The composition is a union of orders, not a reinterpretation of instructions. Two other readings are possible, and both machines reject them (\cref{sec:tso}). In one, a TSO store is a release, since TSO orders it after every earlier access; Arm would then order \ins{STR} before a later \ins{LDAR}. In the other, an annotated access follows Arm alone; \rev{a load could then be reordered with a later \ins{LDAR}}, which Arm allows. Neither holds: a TSO store has only the first order of a release, and an annotation adds Arm's orders without removing TSO's.

In summary, the sequential composition orders every Arm instruction on a TSO thread by the union of the preserved program orders of x86-TSO and Arm, with one addition: every AMO is locked, as an x86 RMW is. On an Arm thread only an AMO with both acquire and release\rvd{, and only if it writes,} is a fence; on a TSO thread every AMO is.

\subsection{Parallel Composition: The Compound Model}
\label{sec:models-par}

The compound model of Arm and TSO keeps each thread's own preserved program order and shares everything else~\cite{heterogen,cmcm-pldi23}. Let $\armppo$ be Arm's \ppo\ restricted to pairs of events on Arm threads, and $\tsoppo$ the sequential composition above restricted to pairs on TSO threads. The compound model has the same three axioms, with
\[
\textsf{acyclic}(\armppo \cup \tsoppo \cup \rfe \cup \coe \cup \fre)
\]
as the order axiom. Sc-per-location and atomicity are unchanged and apply to all threads. The parallel composition is thus built on the sequential one: each TSO thread of the compound model runs Arm instructions under the sequential composition.

\Cref{fig:compound-tests} shows the compound model on mixed versions of the tests of \cref{fig:models-tests}. In IRIW with TSO writers and Arm readers, the readers need acquire loads for the outcome to be forbidden, exactly as when every thread is Arm; with plain Arm readers it is allowed. \rvd{In n6 with a TSO reader and an Arm writer whose second store is a release, the outcome remains allowed: the TSO thread may still read its own write early, and this is the only way to reach the outcome.}

\begin{figure}[!htbp]
\centering
\small
\begin{tabular}{@{}lll@{}}
\toprule
Test & Threads & Compound model\\
\midrule
IRIW & writers \tsow{TSO}, readers \armw{Arm} with acquire loads & forbidden\\
IRIW & writers \tsow{TSO}, readers \armw{Arm} with plain loads & allowed\\
IRIW & writers \armw{Arm}, readers \tsow{TSO} with plain loads & forbidden\\
n6 & \rvd{Thread 0 \tsow{TSO}, Thread 1 \armw{Arm} with a release store} & allowed\\
\bottomrule
\end{tabular}
\caption{The compound model on mixed tests. Verdicts are from herd.}
\label{fig:compound-tests}
\end{figure}

\subsection{Encoding in herd}

Herd ships an Arm model and an x86-TSO model, but they are written in different styles over different instruction sets, and the compound model has to choose a \ppo\ per thread. We therefore re-encode both in the three-axiom form of \cref{sec:axioms}, over AArch64 instructions, so that the models differ only in \ppo\ (\cref{fig:cat}). Arm's \ppo\ (\textsf{arm-ppo.cat}, following~\cite{arm-pulte}) and the sequential composition (\textsf{tso-ppo.cat}) are each written once, as relations over all events. \textsf{tso-ppo.cat} writes the composition literally, as the union of TSO's orders over every load and store, with AMOs locked, and Arm's. Three short files then say which \ppo\ applies where: \textsf{arm.cat} applies Arm's to every thread, \textsf{tso.cat} applies the sequential composition to every thread, and \textsf{compound.cat} applies Arm's to Arm threads and the sequential composition to TSO threads. Tests are written in AArch64 syntax, and a thread whose first instruction is \ins{ISB} runs under TSO.

Because the encoding is our own, we checked it against herd's models. On all 135 tests whose threads all run in Arm mode, \textsf{arm.cat} gives the same verdict as herd's own Arm model (\textsf{aarch64.cat}). The 18 tests whose threads all run in TSO mode and whose instructions are only \ins{LDR}, \ins{STR} and \ins{DMB} are x86 programs; we translate each into x86 syntax (\ins{MOV} for loads and stores, \ins{MFENCE} for \ins{DMB}), and herd's x86-TSO model (\textsf{x86tso.cat}) on the translation gives the same verdict as \textsf{tso.cat} on the original. On tests whose threads all run in one mode, \textsf{compound.cat} gives the same verdict as \textsf{arm.cat} or \textsf{tso.cat}.

\begin{figure}[!htbp]
\begin{lstlisting}[style=cat,firstline=9]
let CASR    = domain([Exp & R]; iico_data+; [PRED])    (* read of a CAS, failed or not *)
let Locked  = M & (domain(amo) | range(amo) | CASR)
let x86-ppo = [M]; (po \ (W * R)); [M]               (* all but W->R *)
            | [M]; po; [DMB-FULL]; po; [M]            (* MFENCE *)
            | [W]; po; [Locked & R]                   (* locked RMW *)
            | [Locked & W]; po; [R]                   (*   acts as a fence *)
let tso-ppo-all = x86-ppo | arm-ppo-all
\end{lstlisting}
\begin{lstlisting}[style=cat,firstline=3]
let obs = rfe | coe | fre                        (* communication *)
acyclic po-loc | rf | co | fr  as sc-per-location
empty rmw & (fre; coe)         as atomicity
acyclic ppo | obs              as order
\end{lstlisting}
\begin{lstlisting}[style=cat,firstline=7,lastline=10]
let MARK = ISB & ~range(po)                   (* a thread starting with ISB runs TSO *)
let TSO  = MARK | range([MARK]; po)
let ARM  = ~TSO
let ppo = [ARM]; arm-ppo-all; [ARM] | [TSO]; tso-ppo-all; [TSO]
\end{lstlisting}
\caption{The herd encoding of the proposal\rvc{, against which every verdict in \cref{sec:arm,sec:tso,sec:compound} is computed. Top: the sequential composition (\textsf{tso-ppo.cat}), x86-TSO's order with every AMO and every \ins{CAS} read locked, united with Arm's. Middle: the three axioms of \cref{sec:axioms}, shared by every model (\textsf{axioms.cat}). Bottom: the compound model (\textsf{compound.cat}), which applies Arm's order to Arm threads and the sequential composition to TSO threads. \Cref{tab:seq} states the same orders in prose; the listing is the exact definition, for readers who want to check it or run herd.} Arm's preserved program order (\textsf{arm-ppo.cat}) follows~\cite{arm-pulte}; all model files are in the artifact (\cref{sec:artifacts}).}
\label{fig:cat}
\end{figure}

\subsection{The Apple Models}
\label{sec:models-apple}

The experiments of \cref{sec:arm,sec:tso,sec:compound} find that both machines preserve three orders that neither Arm nor the sequential composition requires. We define them here, so that all the models are in one place; the evidence for each is in those sections. We keep an order only if both machines enforce it, and add each to every thread, Arm or TSO:
\begin{itemize}
\item[A1] a store and a later acquire load (\ins{LDAPR}, \ins{LDAR}, \ins{LDAXR}) of the same address are ordered: an acquire load never reads its own thread's store early\rev{ (\ins{LDAXR} tested on TSO threads only)};
\item[A2] \ins{DMB LD} and \ins{DSB LD} order a store before every later access;
\item[A3] a load is ordered before a later store.
\end{itemize}
Arm already orders a load before every access after a \ins{DMB LD}, so with A2 a \ins{DMB LD} is a full barrier. A3 adds nothing on a TSO thread, where TSO already preserves load $\to$ store; it matters only on Arm threads. The Apple models also treat a \ins{CAS} as an AMO that always writes: one whose comparison fails writes back the value it read (\cref{sec:arm-atomics}). This too adds nothing on a TSO thread, where a failing \ins{CAS} is already locked. \textsf{arm-apple.cat}, \textsf{tso-apple.cat} and \textsf{compound-apple.cat} are \textsf{arm.cat}, \textsf{tso.cat} and \textsf{compound.cat} with the three added to each thread's \ppo\ and a failing \ins{CAS} writing back; nothing else changes (\cref{fig:cat-apple}). On tests whose threads all run in one mode, \textsf{compound-apple.cat} gives the same verdict as \textsf{arm-apple.cat} or \textsf{tso-apple.cat}. The first three models are the proposal; the last three describe the two machines.

\begin{figure}[!htbp]
\begin{lstlisting}[style=cat,firstline=6]
let apple-all = [W]; po-loc; [R & (A | Q)]            (* A1 *)
              | [W]; po; [DMB-LD | DSB-LD]; po; [M]   (* A2 *)
              | [R]; po; [W]                          (* A3 *)

(* A CAS is an AMO that always writes:
   one whose comparison fails writes back the value it read. *)
let CASR = domain([Exp & R]; iico_data+; [PRED])
empty CASR \ domain(amo) as cas-writes
\end{lstlisting}
\caption{The orders both machines preserve beyond the models (\textsf{apple-ppo.cat}). The Apple models add \textsf{apple-all} to each thread's \ppo, and add to the axioms the rule \textsf{cas-writes}, which says that a \ins{CAS} writes even when its comparison fails, writing back the value it read.}
\label{fig:cat-apple}
\end{figure}

%% file: sections/method.tex
\section{Methodology}
\label{sec:method}

\paragraph{The machines.} We test two machines. The first is an Apple Mac mini (M1, 2020) in the GCC Compile Farm, running Debian with an Asahi Linux 6.5 kernel. The M1 has four performance cores (Firestorm, up to 3.2\,GHz) and four efficiency cores (Icestorm, up to 2.06\,GHz), and 8\,GB of memory. The second is an Apple MacBook Air (M2), running Fedora Asahi Linux with a 7.2 kernel. The M2 has four performance cores (Avalanche) and four efficiency cores (Blizzard). \newtext{\rev{For the long runs of \cref{sec:arm-atomics,sec:scale}} we also use macOS 13.5, on the M2 and on a Mac Studio (M2 Max, 2023). macOS offers no call to set a thread's mode or to pin a thread, so under macOS we run only tests whose threads are all in Arm mode, with threads unpinned.}

\paragraph{The TSO bit.} Apple cores have a control bit that switches the core's memory model to TSO. Linux exposes it per thread through a \ins{prctl} call, \ins{PR\_SET\_MEM\_MODEL}, and saves and restores it when it switches threads. Each test thread sets its mode as its first action and reads it back with \ins{PR\_GET\_MEM\_MODEL}; a mismatch aborts the run. \rve{As a check that the bit takes effect, message passing with plain accesses never appears between two TSO threads, but appears as soon as either thread is in Arm mode.}

\subsection{Choosing the Tests}
\label{sec:method-tests}

The models of \cref{sec:background,sec:compose} differ only in each thread's preserved program order, and share three axioms (\cref{sec:axioms}). The tests ask three questions. Does each thread keep the preserved program order of its mode? Do writes propagate as the order axiom says? Are RMWs atomic, and which are locked?

\paragraph{Preserved program order.} We expand the preserved program order of Arm and of the sequential composition into a table. A row is a mode, a kind of pair (store $\to$ store, load $\to$ load, load $\to$ store or store $\to$ load) and a mechanism: the forms of the two accesses (\ins{LDR}, \ins{LDAPR}, \ins{LDAR}; \ins{STR}, \ins{STLR}), a barrier between plain accesses (\ins{DMB}, \ins{DSB} or their \ins{LD} and \ins{ST} forms), or a dependency (address, data, control, or control and \ins{ISB}). Both modes run the same Arm instructions, so each has the same 56 mechanisms, 112 rows in all. In Arm mode their verdicts are those of the Arm column of \cref{tab:ppo}; in TSO mode, those of \cref{tab:seq}. Each row is tested in the two-thread test in which its pair is one thread's two accesses: MP for store $\to$ store (the writer) and load $\to$ load (the reader), LB for load $\to$ store, and SB for store $\to$ load. The row's pair is one edge of the cycle, and the other thread's edge gets a mechanism its model orders for certain: a full \ins{DMB} on an Arm thread; on a TSO thread nothing, since TSO already orders it, except in SB, where that edge is a store followed by a load and gets a \ins{DMB}. The outcome is thus forbidden exactly when the row orders its pair. The other thread runs once in each mode.

\paragraph{Propagation.} The order axiom fixes how writes propagate between threads, and three properties follow from it. First, a write seen by one other thread is seen by all, the first half of other-multi-copy atomicity (\cref{sec:axioms}); WRC, RWC and IRIW test it. Second, order carries along a chain of threads, which Power calls cumulativity; ISA2, message passing through three threads, tests it. We run each of the four shapes under every assignment of modes to its threads, including threads that only write. Each edge is ordered by one of the \emph{doubtful mechanisms}, the weakest that should order it, and every combination is run. On an Arm thread the doubtful mechanisms are a dependency, \ins{LDAPR}, \ins{STLR} or \ins{DMB ST}, and for a store followed by a load, \ins{STLR} then \ins{LDAR}. On a TSO thread they are TSO's own order or the annotations \ins{LDAPR} and \ins{STLR}, which x86 does not have, and for a store followed by a load, \ins{DMB} (as \ins{MFENCE}) or \ins{STLR} then \ins{LDAR}. Full barriers are not among the doubtful mechanisms: they are built to order other threads' writes, so a weaker mechanism would fail first. Each edge also has a control, in which that edge is plain and the other edges are ordered.

Third, a thread may read its own store early, the other half of other-multi-copy atomicity, as x86-TSO allows and the multi-copy-atomic IBM 370 does not. We test it with n6~\cite{x86tso-cacm} and n7~\cite{x86tso-tphols}, written by hand, in which a thread reads back its own store. We vary the store (\ins{STR}, \ins{STLR}, \ins{STXR}) and the load that reads it back (\ins{LDR}, \ins{LDAPR}, \ins{LDAR}), and run the reading thread in Arm mode and in TSO mode.

\paragraph{RMWs.} Atomicity and locking act on a store followed by a load, so the RMW tests are Dekker tests: SB with RMWs. On a TSO thread we test every store $\to$ load pair in which either access is an exclusive or an AMO, 100 pairs in all, each in SB as the store $\to$ load rows are. In six of them a store precedes an LL/SC pair that no release orders; atomicity, not preserved program order, forbids their outcome (\cref{sec:models-seq}). In each mode, hand-written Dekker tests replace each thread's load, its store, or neither by an RMW, the last as a candidate fence, in each AMO and LL/SC form.

\paragraph{Controls.} A forbidden outcome that never appears is evidence only if the test could have shown it. Each generated forbidden test therefore has a \emph{control}, an allowed test that differs from it only at one pair: that pair left plain or, where TSO orders the pair whatever the mechanism, that thread in Arm mode. A machine enforces the pair if the forbidden outcome never appears and the control's does.

\paragraph{Coverage and verdicts.} A script checks coverage: every row and every RMW pair has a test with a partner of each mode, every combination of the doubtful mechanisms runs under every assignment of modes, every generated forbidden test names an allowed control, and herd's verdict for every test agrees with its row. In all there are 936 tests, 658 generated and 278 written by hand. herd gives each the verdict of the Arm model, the sequential composition, the compound model and herd's own Arm model, and, for the x86 programs, whose TSO threads use only plain accesses and \ins{DMB}, of herd's x86-TSO model, after translating them to x86. The results are reported by mode: \cref{sec:arm} all Arm, \cref{sec:tso} all TSO, and \cref{sec:compound} mixed.

\subsection{Running the Tests}

We run every test with litmus7~\cite{litmus7}, unmodified except for the per-thread mode call, in six configurations: two synchronization modes of litmus7, \textsf{user} and \textsf{timebase}, times three placements of the threads, all on performance cores, all on efficiency cores, and spread across both clusters. Each configuration runs $10^8$ iterations, so every test runs $6\times10^8$ times on each machine. No configuration is redundant: \rve{most outcomes seen ten times or fewer appeared in only one of them, and which one differs from shape to shape}. For LL/SC pairs we count only the iterations in which every store-exclusive succeeds. One difference in method concerns the M2. When a test's threads may run on both clusters, the Linux scheduler on the M2 placed two of litmus7's spinning threads on one core, which slowed those runs by two to three orders of magnitude. On the M2 we therefore pin each thread of such a run to the least loaded core of the run's set, breaking ties at random, so that which thread runs on which kind of core varies between runs. Runs on a single cluster are not pinned, and the M1 runs are not pinned. One script runs the whole set on either machine.

\subsection{Reading the Results}
\label{sec:method-reading}

The rarest outcomes we observe occur about once in $6\times10^8$ runs, so a zero is not proof that an outcome is forbidden. A zero for a forbidden outcome therefore counts only if its control was observed on the same machine (for a hand-written test, the closest allowed test of its shape); otherwise we report the test as not exercised. The exception is load buffering, which neither machine shows in any form. Its tests have no control, so we report their outcomes as forbidden by A3 (\cref{sec:models-apple}), which rests not on a control but on how cores are typically built: they commit instructions in order and make a store visible only after it commits, so a store cannot be seen before an earlier load. \newtext{In runs 40 times longer, load buffering still never appeared, although in millions of iterations its two threads ran their accesses at the same time, \rev{as shown by both loads reading 0} (\cref{sec:scale}).} A control that never appears is either explained by an order the machines add or reported as rare. Rare counts vary severalfold between runs of the same test, so the tables give only orders of magnitude.

%% file: sections/results-arm.tex
\section{Arm Mode}
\label{sec:arm}

Does Arm mode follow the Arm model? It does: on neither machine was an outcome the Arm model forbids observed, in any of the 135 tests whose threads all run in Arm mode. Both machines are slightly stronger than the model, in the three orders A1--A3 of \cref{sec:models-apple} and in how a failing \ins{CAS} behaves. We take the three questions of \cref{sec:method-tests} in turn: preserved program order, propagation, and RMWs.

\subsection{Preserved Program Order}

\input{tables/res-arm-ppo}

\Cref{tab:arm-results} gives every Arm row of preserved program order, each in its two-thread test (\cref{sec:method-tests}) with the other thread also in Arm mode. In every row that Arm orders, the outcome never appeared on either machine. A zero is evidence only if the row's control (last column) appears (\cref{sec:method-reading}). The rows that Arm does not order show how often each kind of pair is reordered. A store and a later load are reordered $10^5$ to $10^7$ times in $6\times10^8$. Two loads are reordered \audit{tens to hundreds of times} on the M1 and at most tens of times on the M2. Two stores are reordered hundreds of times on the M1 and at most tens of times on the M2 (\cref{sec:compound-partner}). A load and a later store are never reordered, in any row of either mode; this is why we add A3, a load ordered before a later store. Load buffering thus has no control, and that it never appears rests on the microarchitectural argument of \cref{sec:method-reading}: cores typically commit instructions in order and make a store visible only after it commits, so a store cannot be seen before an earlier load.

\paragraph{A barrier that orders loads only.} Arm's \ins{DMB LD} and \ins{DSB LD} order earlier loads before later accesses, and nothing else. With either barrier between a store and a later load, the outcome never appears, while every other unordered store $\to$ load row appears $10^5$ times or more. We therefore add A2: these barriers also order a store before every later access, which makes them full barriers. With either barrier between two stores the outcome never appears either, as A2 predicts.

\paragraph{Reading a thread's own store.} The last rows of \cref{tab:arm-results} are n7 tests of Owens et al.~\cite{x86tso-tphols}: a thread stores \ins{x}, reads it back, and then reads \ins{y}, with the two loads ordered by an acquire or an address dependency; an observer reads \ins{y} and then \ins{x}. The outcome appears only if the read-back takes its value from the thread's own store before the store is visible to other threads, that is, if the load reads its own store early. Arm forbids this only for a release store followed by an \ins{LDAR}, and allows it for every other pair. With a plain load, both machines show it, whether the store is plain, a release or a store-exclusive, $10^3$ to $10^5$ times. With an acquire load, \ins{LDAPR} or \ins{LDAR}, they never do, whatever the store. We therefore add A1: a store is ordered before a later acquire load of the same address, so that an acquire load never reads its own thread's store early.

\paragraph{Zeros we do not explain.} Two unordered rows never appear and A1--A3 do not account for them: \ins{STLR} followed by \ins{STR}, and \ins{DSB ST} between two loads. Both are pairs the machines rarely reorder even when they are plain: two plain stores at most hundreds of times on the M1 and at most tens on the M2, and two plain loads hundreds of times on the M1 and at most tens on the M2. A zero in a row whose plain form is this rare could be chance, so we do not add an order for either row.

\subsection{Three and Four Threads}
\label{sec:arm-multi}

\input{tables/res-arm-multi}

\Cref{tab:arm-multi} gives WRC, ISA2, RWC and IRIW with every thread in Arm mode, under every combination of the doubtful mechanisms: address and data dependencies, \ins{LDAPR}, \ins{STLR} and \ins{DMB ST} (\cref{sec:method-tests}). None of the forbidden outcomes was observed. RWC and IRIW are well exercised: with one edge left unordered, the outcome appears $10^2$ to $10^5$ times. WRC and ISA2 are exercised far less: their controls appeared only a few times, WRC only on the M1 and ISA2 only on the M2.

\subsection{RMWs}
\label{sec:arm-atomics}

\input{tables/res-arm-rmw}

Recall from \cref{sec:rmw} and \cref{fig:rmw-types} the two orders atomicity gives an RMW: Arm's type-3 atomicity orders a store before a later release RMW's read, and type-1 atomicity would also order an acquire RMW's write before later accesses, which Arm states only for AMOs with both acquire and release. We test RMWs in three forms of Dekker (\cref{tab:arm-atomics}): the RMW replaces each thread's load, or its store, or it accesses a third location between the store and the load, as a candidate fence.

\paragraph{The release case.} With a release RMW in place of each load, as in \cref{fig:rmw-types}(a), the outcome is forbidden by the Arm model through atomicity (\cref{sec:rmw}). Removing the atomicity axiom from herd's Arm model makes the outcome allowed, for both \ins{SWPL} and \ins{LDXR}/\ins{STLXR}. For the LL/SC pair this holds only when both store-exclusives succeed; if one fails it writes nothing, and Arm allows the outcome. Both machines respect the order for both kinds of RMW, in the runs in which both store-exclusives succeed (\cref{tab:arm-atomics}).

\paragraph{The acquire case.} Now let an acquire RMW replace each thread's store, as in \cref{fig:rmw-types}(b). The Arm model allows the outcome for both kinds of RMW, and a type-1 RMW would forbid it (\cref{sec:rmw}). With the release AMOs \ins{SWPL}, \ins{LDADDL} or \ins{CASL}, or the acquire LL/SC pair \ins{LDAXR}/\ins{STXR}, the outcome appears about $10^7$ times on both machines. With the acquire AMOs \ins{SWPA}, \ins{LDADDA} or \ins{CASA} it never appears in $6\times10^8$ iterations, which suggests an order. But the outcome is rare, not forbidden. In $10^{11}$ iterations on the M2 with unpinned threads, \ins{SWPA} showed it 34 times under macOS and 4 times under Linux, and under macOS \ins{LDADDA} and \ins{CASA} showed it in $10^{10}$ iterations each. It also appears on the M1 and the M2 Max (\cref{sec:scale}). The machines therefore do not order the write of an acquire AMO before later accesses, and we add no order for it.

\paragraph{RMWs as fences.} An acquire AMO is not a fence either. Between a store and a load, \ins{SWPA} shows the outcome $10^5$ times on both machines, and the release AMO and every LL/SC pair $10^4$ to $10^6$ times; only \ins{SWPAL}, which has both orderings, prevents it, as Arm specifies.

\paragraph{A failing CAS.} A \ins{CAS} whose comparison fails need not write (\cref{sec:rmw}), and the Arm model then does not order an earlier store before it, so it allows Dekker with a failing \ins{CASAL} between the store and the load. Neither machine shows the outcome, while a failing \ins{CAS}, \ins{CASA} or \ins{CASL} in the same place shows it $10^4$ to $10^5$ times on the M1 and $10^{4}$ times on the M2. The machines behave as if a \ins{CAS} were an AMO like any other, writing back the value it read when its comparison fails; Arm's own rule then makes a failing \ins{CASAL} a full fence.

\paragraph{An allowed outcome never observed.} Dekker with \ins{SWPA} in place of the loads is allowed even with A1--A3, and neither machine shows it. We do not add an order to forbid it. The outcome needs the store to become visible after the RMW's write, two stores reordered, which is rare on both machines; with the acquire LL/SC pair in place of the loads it appears hundreds of times on the M1 and tens of times on the M2.

\paragraph{Summary.} Arm mode follows the Arm model. Both machines are stronger in three ways: an acquire load never reads its own thread's store early (A1); \ins{DMB LD} and \ins{DSB LD} are full barriers (A2); and a load is never reordered with a later store (A3). A failing \ins{CAS} also behaves as if it writes, which makes a failing \ins{CASAL} a fence. A1, A2 and the failing \ins{CASAL} rest on zeros next to observed neighbours; A3 has no control, since no test in either mode shows load buffering. \newtext{Run again at \rev{up to} 40 times the iterations, all four zeros held (\cref{sec:scale}). A fifth zero did not: Dekker with an acquire AMO in place of the store is rare rather than forbidden, and we add no order for it.}

%% file: tables/res-arm-ppo.tex
\begin{table}[!htbp]
\centering
\footnotesize
\setlength{\tabcolsep}{3pt}
\begin{tabularx}{\linewidth}{@{}>{\raggedright\arraybackslash}p{0.11\linewidth}>{\raggedright\arraybackslash}Xllll>{\raggedright\arraybackslash}p{0.24\linewidth}@{}}
\toprule
Pair & Mechanism & Shape & \armw{Arm} & M1 & M2 & Control: M1 / M2\\
\midrule
store $\to$ store & \ins{STR};\ins{STLR}, \ins{STLR};\ins{STLR}, \ins{DMB}, \ins{DMB ST}, \ins{DSB}, \ins{DSB ST} (6) & MP & ordered & \rfz & \rfz & plain: \rsn{$10^{2}$} / \rux; \tsow{TSO} partner: \rsn{$10^{1}$} / \rsn{$10^{1}$}\\
 & plain & MP & not ordered & \rsn{$10^{2}$} & \rux & \\
 & \ins{STLR};\ins{STR} & MP & not ordered & \rux & \rux & \\
 & \ins{DMB LD}, \ins{DSB LD} & MP & not ordered & \rst{A2} & \rst{A2} & \\
\addlinespace
load $\to$ load & acquire first, \ins{DMB}, \ins{DMB LD}, \ins{DSB}, \ins{DSB LD}, address dependency, control dependency and \ins{ISB} (12) & MP & ordered & \rfz & \rfz & plain: \rsn{$10^{2}$} / \rsn{$10^{0}$}\\
 & plain; \ins{LDR};\ins{LDAPR}; \ins{LDR};\ins{LDAR}; \ins{DMB ST} & MP & not ordered & \rsn{$10^{2}$} & \rsn{$10^{0}$--$10^{1}$} & \\
 & control dependency & MP & not ordered & \rsn{$10^{1}$} & \rsn{$10^{1}$} & \\
 & \ins{DSB ST} & MP & not ordered & \rux & \rux & \\
\addlinespace
load $\to$ store & release or acquire, \ins{DMB}, \ins{DMB LD}, \ins{DSB}, \ins{DSB LD}, dependencies (13) & LB & ordered & \rfz & \rfz & plain: \rst{A3} / \rst{A3}\\
 & plain, \ins{DMB ST}, \ins{DSB ST} & LB & not ordered & \rst{A3} & \rst{A3} & \\
\addlinespace
store $\to$ load & \ins{STLR};\ins{LDAR}, \ins{DMB}, \ins{DSB}, \ins{DSB ST} (4) & SB & ordered & \rfz & \rfz & plain: \rsn{$10^{6}$} / \rsn{$10^{5}$}\\
 & plain; \ins{STR};\ins{LDAPR}; \ins{STR};\ins{LDAR}; \ins{DMB ST} & SB & not ordered & \rsn{$10^{6}$} & \rsn{$10^{5}$--$10^{6}$} & \\
 & \ins{STLR};\ins{LDR}; \ins{STLR};\ins{LDAPR} & SB & not ordered & \rsn{$10^{6}$--$10^{7}$} & \rsn{$10^{5}$--$10^{6}$} & \\
 & \ins{DMB LD}, \ins{DSB LD} & SB & not ordered & \rst{A2} & \rst{A2} & \\
\addlinespace
store $\to$ read-back & \ins{STLR};\ins{LDAR} & n7 & ordered & \rfz & \rfz & \ins{STLR};\ins{LDR}: \rsn{$10^{3}$} / \rsn{$10^{4}$}\\
 & \ins{STR};\ins{LDR} & n7 & not ordered & \rsn{$10^{4}$} & \rsn{$10^{4}$} & \\
 & \ins{STLR};\ins{LDR} & n7 & not ordered & \rsn{$10^{3}$} & \rsn{$10^{4}$} & \\
 & \ins{STXR};\ins{LDR} & n7 & not ordered & \rsn{$10^{3}$--$10^{4}$} & \rsn{$10^{5}$} & \\
 & \ins{STR}, \ins{STLR} or \ins{STXR}, then \ins{LDAPR} or \ins{LDAR} & n7 & not ordered & \rst{A1} & \rst{A1} & \\
\bottomrule
\end{tabularx}
\caption{The Arm rows of preserved program order, every thread in Arm mode, each test run $6\times10^8$ times on each machine. An entry is the order of magnitude of the number of times the outcome appeared ($10^k$: from $10^k$ to $10^{k+1}-1$; a range spans the row's tests). An ordered row's control is the same test without its ordering: plain accesses, with no barrier or dependency; for n7, \ins{LDR} in place of \ins{LDAR}. The store $\to$ store control is also given with a TSO-mode partner, the only partner with which it appeared on the M2 (\cref{sec:compound-partner}). Shading: \rfz\ and \rsn{$10^{4}$}, the machines agree with the model; \rst{A2}, allowed by the model but never seen, and forbidden by A2 (\cref{sec:models-apple}); \rux, allowed but never seen, and no order of the machines forbids it.}
\label{tab:arm-results}
\end{table}

%% file: tables/res-arm-multi.tex
\begin{table}[!htbp]
\centering
\small
\begin{tabular}{@{}lrrrll@{}}
\toprule
& \multicolumn{2}{c}{Forbidden} & \multicolumn{3}{c}{Allowed controls}\\
\cmidrule(lr){2-3}\cmidrule(lr){4-6}
Shape & tests & seen & tests & M1 & M2\\
\midrule
RWC & 2 & \rfz & 2 & \rsn{$10^{4}$--$10^{5}$} & \rsn{$10^{4}$--$10^{5}$}\\
IRIW & 4 & \rfz & 2 & \rsn{$10^{2}$--$10^{3}$} & \rsn{$10^{3}$--$10^{4}$}\\
WRC & 5 & \rfz & 1 & \rsn{$10^{0}$} & \rux\\
ISA2 & 9 & \rfz & 2 & \rux & \rsn{0--$10^{0}$}\\
\bottomrule
\end{tabular}
\caption{WRC, ISA2, RWC and IRIW with every thread in Arm mode, under every combination of the doubtful mechanisms (\cref{sec:method-tests}), $6\times10^8$ iterations of each test on each machine. Forbidden by the Arm model with A1--A3. A control is a forbidden test with one edge left unordered; controls that A3 forbids are omitted, and the range spans the shape's controls. Entries as in \cref{tab:arm-results}.}
\label{tab:arm-multi}
\end{table}

%% file: tables/res-arm-rmw.tex
\begin{table}[!htbp]
\centering
\footnotesize
\setlength{\tabcolsep}{4pt}
\begin{tabular}{@{}llllllll@{}}
\toprule
& & \multicolumn{3}{c}{AMO} & \multicolumn{3}{c}{LL/SC pair}\\
\cmidrule(lr){3-5}\cmidrule(lr){6-8}
Dekker: the RMW & Ordering & \armw{Arm} & M1 & M2 & \armw{Arm} & M1 & M2\\
\midrule
replaces the load & release & forbidden & \rfz & \rfz & forbidden & \rfz & \rfz\\
 & acquire & allowed & \rux & \rux & allowed & \rsn{$10^{2}$} & \rsn{$10^{1}$}\\
\addlinespace
replaces the store & release & allowed & \rsn{$10^{7}$} & \rsn{$10^{7}$} & allowed & \rsn{$10^{7}$} & \rsn{$10^{7}$}\\
 & acquire & allowed & \rux & \rux & allowed & \rsn{$10^{7}$} & \rsn{$10^{7}$}\\
 & both & forbidden & \rfz & \rfz & allowed & \rsn{$10^{7}$} & \rsn{$10^{7}$}\\
\addlinespace
as a fence & release & allowed & \rsn{$10^{6}$} & \rsn{$10^{5}$} & allowed & \rsn{$10^{5}$} & \rsn{$10^{4}$}\\
 & acquire & allowed & \rsn{$10^{5}$} & \rsn{$10^{5}$} & allowed & \rsn{$10^{5}$} & \rsn{$10^{4}$}\\
 & both & forbidden & \rfz & \rfz & allowed & \rsn{$10^{5}$} & \rsn{$10^{4}$}\\
\bottomrule
\end{tabular}
\caption{Dekker with RMWs, both threads in Arm mode: the RMW replaces each thread's load (\ins{STR x}; RMW \ins{y}), its store (RMW \ins{x}; \ins{LDR y}), or a third location between the two (\ins{STR x}; RMW \ins{z}; \ins{LDR y}). The AMO is \ins{SWP}; where the RMW replaces the store, also \ins{LDADD} and \ins{CAS}, whose rows agree. The LL/SC pair is \ins{LDXR}/\ins{STXR}, with \ins{LDAXR} for acquire and \ins{STLXR} for release. For LL/SC pairs only runs in which both store-exclusives succeed are counted, $4.4\times10^8$ to $6\times10^8$. Entries as in \cref{tab:arm-results}.}
\label{tab:arm-atomics}
\end{table}

%% file: sections/results-tso.tex
\section{TSO Mode}
\label{sec:tso}

We ask two questions of TSO mode. The first concerns programs that use only what x86 also has, plain loads and stores and the full \ins{DMB}: does TSO mode implement x86-TSO, in which a load may read its own thread's store before other threads see it, or a stronger model such as the IBM 370's, in which it may not? The second concerns the Arm instructions whose meaning goes beyond x86's, acquire loads, release stores, the other barriers and the RMWs: does a TSO thread follow the sequential composition (\cref{sec:models-seq})? 

TSO mode implements x86-TSO: a load reads its own thread's store early, and often. With the other Arm instructions it follows the sequential composition: on neither machine was an outcome the sequential composition forbids observed, in any of the 272 tests whose threads all run in TSO mode. TSO mode is stronger than the sequential composition in two of the orders Arm mode adds, A1 and A2 (\cref{sec:models-apple}); the third, A3, orders a load before a later store, which TSO already does. As in \cref{sec:arm}, we take the three questions of \cref{sec:method-tests} in turn: preserved program order, propagation, and RMWs.

\subsection{Preserved Program Order}
\label{sec:tso-seq}

\input{tables/res-tso-ppo}

\Cref{tab:tsoarm-results} gives every row of the sequential composition, with both threads in TSO mode, followed by the read-back rows. In every row that the sequential composition orders, the outcome never appeared on either machine. As in \cref{sec:arm}, a zero is evidence only if the row's control (last column) appears.

\paragraph{TSO's orders are kept.} The sequential composition keeps TSO's order on every store $\to$ store, load $\to$ load and load $\to$ store pair, whatever the annotation or barrier\rev{, for example a load before a later \ins{LDAR}}. None of these 44 rows was observed. Their control is the same test with that thread in Arm mode, where Arm does not order the pair, and the controls are as uneven as in Arm mode: an Arm-mode reader reorders two loads $10^2$ to $10^3$ times on the M1 and $10$ to $10^3$ times on the M2, and an Arm-mode writer reorders two stores \rev{tens of times} on each machine. Load $\to$ store has no control, since no thread in either mode reorders a load with a later store (\cref{sec:arm}). \rev{Four other rows have no observed control: \ins{STLR} before \ins{STR}, \ins{DMB LD} or \ins{DSB LD} between two stores, and \ins{DSB ST} between two loads. Where the control appears,} an Arm annotation or barrier on a TSO thread never weakens TSO's order.

\paragraph{Store $\to$ load.} The sequential composition adds Arm's orders to TSO only on store $\to$ load, and there only a release store followed by an \ins{LDAR}, and the full barriers, order the pair. The machines agree. The four ordered rows were never observed, and every unordered row but two shows the outcome $10^4$ to $10^6$ times. In particular, a plain store is not a release here, although TSO orders it after earlier accesses as it would a release: \ins{STR} followed by \ins{LDAR} shows the outcome \audit{$10^5$ to $10^6$ times, as} two plain accesses do. The two exceptions are \ins{DMB LD} and \ins{DSB LD}. They have no x86 counterpart, and on Arm they order earlier loads only, so the rule allows the outcome with them; neither machine shows it. This is the same strengthening as in Arm mode, A2: on both machines these barriers are full barriers in either mode.

\paragraph{x86's operations.} Plain loads and stores, and the full \ins{DMB}, which is \ins{MFENCE}, are x86's own operations. Eighteen of the tests use only these on every thread, so they are x86 programs; among them are MP, SB, LB, WRC, ISA2, RWC, IRIW, n6 and n7. We translated each to x86, and herd's x86-TSO model gives each the verdict of the sequential composition. Both machines agree with it on every one: the twelve outcomes it forbids never appear, and the six it allows appear on both machines.

\paragraph{Reading a thread's own store.} The read-back rows answer the first question. In n7~\cite{x86tso-tphols} a thread stores \ins{x}, reads it back and then reads \ins{y}, while an observer reads \ins{y} and then \ins{x}. x86-TSO allows the outcome, because the thread may read \ins{x} from its store buffer before the store is visible to other threads; a TSO without that rule, such as the IBM 370's, forbids it. With every thread in TSO mode and plain accesses, it appears $10^4$ times on the M1 and $10^3$ on the M2. So does n6~\cite{x86tso-cacm}, the two-thread test of the same question ($10^5$ and $10^4$ times), while n6 with an \ins{MFENCE} never does. TSO mode thus implements x86-TSO, not a stronger TSO.

\paragraph{Which loads read their own store early.} The second question asks the same of the Arm instructions: on a TSO thread, which loads may read their own thread's store before other threads see it? We ran n7 with every combination of a store form and a load form for the store and the read-back (\cref{tab:tsoarm-results}). The sequential composition lets every load read early except in two cases: \ins{LDAR} or \ins{LDAXR} after a release store, and any load after an AMO. Neither case was observed.

A plain load reads early after a plain or release store, $10^3$ to $10^4$ times. An acquire load, \ins{LDAPR}, \ins{LDAR} or \ins{LDAXR}, never reads early, after any store, on either machine, while the same test with a plain load does. The sequential composition does not require this; it is A1, the order the machines add in Arm mode. A load-exclusive reads early only rarely, at most tens of times. It never does after \ins{STLXR}, but with an outcome this rare, that zero is not evidence of an order.

\paragraph{A release store-exclusive.} One row differs between the machines. On the M1, a plain load never reads its own store early after \ins{STLXR}, while it does after \ins{STXR}, about ten times. On the M1 alone we would have added this order. The M2 shows the outcome, 78 times in $6\times10^8$ iterations, more often than after \ins{STXR}, so we do not. This is why the model of the machines keeps only the orders both machines enforce.

\subsection{Three and Four Threads}
\label{sec:tso-multi}

\input{tables/res-tso-multi}

\Cref{tab:tso-multi} gives WRC, ISA2, RWC and IRIW with every thread in TSO mode, under every combination of the doubtful mechanisms of a TSO thread: TSO's own order and the annotations \ins{LDAPR} and \ins{STLR}, which x86 does not have (\cref{sec:method-tests}). None of the forbidden outcomes was observed: TSO mode keeps x86-TSO's propagation, also when its accesses carry Arm's annotations. TSO orders every edge of these shapes but store $\to$ load, so a control puts one thread in Arm mode, with plain accesses, except for one control of RWC. RWC and IRIW are well exercised, $10$ to $10^4$ times. WRC is well exercised on the M1 ($10^4$ times) and less on the M2 (tens of times). ISA2 is barely exercised: its controls appear between one and twelve times.

\subsection{RMWs}
\label{sec:tso-rmw}

\input{tables/res-tso-rmw}

The sequential composition treats the two kinds of Arm RMW differently. Every AMO is locked, as an x86 RMW is: it is a type-1 RMW (\cref{sec:rmw}), and with TSO's order this makes it a full fence. An LL/SC pair is a TSO load followed by a TSO store, with Arm's type-3 atomicity, and is not locked. Both machines behave this way (\cref{tab:tso-atomics}).

\paragraph{AMOs.} No test with an AMO shows the outcome, whether the AMO takes the place of the store, takes the place of the load, or lies between them, for \ins{SWP}, \ins{LDADD} and \ins{CAS} with each of the four orderings. The reason is the same in every case. The AMO is locked, so a store before it is ordered before its read, and its write is ordered before a later load; with TSO's other orders, the AMO is a full fence. A \ins{CAS} whose comparison fails is no exception: as \cref{sec:models-seq} explains, it is locked whether or not it writes, and neither machine shows the outcome with one.

\paragraph{LL/SC pairs.} An LL/SC pair orders much less. A store followed by an LL/SC pair is ordered by atomicity, as a store followed by a release RMW is in Arm mode (\cref{fig:rmw-types}(a)); because TSO orders every store before a later store, this holds for every LL/SC pair, not only a release one. It holds only when the store-exclusive succeeds: one that fails writes nothing, and the outcome then appears $10^6$ times on both machines, so the table counts only runs in which the store-exclusives succeed. The rule that orders \ins{STLR} before \ins{LDAR} (\cref{sec:tso-seq}) also orders \ins{STLXR} before \ins{LDAR} or \ins{LDAXR}, and \ins{STLR} before \ins{LDAXR}. None of these pairs shows the outcome.
Otherwise an LL/SC pair orders nothing. Followed by a load, or between a store and a load, it shows the outcome $10^4$ to $10^6$ times.

\paragraph{What this means.} TSO mode thus has an RMW that x86 does not. An x86 RMW is locked, and so is a fence. An LL/SC pair on a TSO thread is atomic, but a later load may read memory before the pair's store-exclusive, or an earlier store, is visible to other threads.

\paragraph{Summary.} \Cref{tab:tso-answers} summarizes how each Arm instruction behaves on a TSO thread. Both machines follow the sequential composition, except in two places where both are stronger: A1 and A2.

\begin{table}[!htbp]
\centering
\footnotesize
\begin{tabular}{@{}lll@{}}
\toprule
Instructions on a \tsow{TSO} thread & Behaviour & Beyond the rule\\
\midrule
\ins{LDR}, \ins{STR}, \ins{DMB SY} & x86 load, store, \ins{MFENCE} & \\
\ins{LDAPR} & TSO load & \exbox{not read early (A1)}\\
\ins{LDAR} & TSO load; ordered after a release store & \exbox{not read early (A1)}\\
\ins{STLR} & TSO store; ordered before \ins{LDAR} & \\
\ins{LDXR}/\ins{STXR}, any annotation & TSO load and store, atomic, not locked & \\
\ins{SWP}, \ins{LDADD}, \ins{CAS} (any form, or failing) & locked RMW, a fence & \\
\ins{DSB SY}, \ins{DSB ST} & full fence & \\
\ins{DMB ST} & no store $\to$ load order & \\
\ins{DMB LD}, \ins{DSB LD} & no store $\to$ load order & \exbox{orders store $\to$ load (A2)}\\
\bottomrule
\end{tabular}
\caption{Arm instructions on a TSO thread. The middle column is what the sequential composition of x86-TSO and Arm gives, and what TSO mode does on both machines; the last column, shaded, is where both machines are stronger.}
\label{tab:tso-answers}
\end{table}

%% file: tables/res-tso-ppo.tex
\begin{table}[!htbp]
\centering
\footnotesize
\setlength{\tabcolsep}{3pt}
\begin{tabularx}{\linewidth}{@{}>{\raggedright\arraybackslash}p{0.11\linewidth}>{\raggedright\arraybackslash}Xllll>{\raggedright\arraybackslash}p{0.24\linewidth}@{}}
\toprule
Pair & Mechanism & Shape & \tsow{TSO} & M1 & M2 & Control: M1 / M2\\
\midrule
store $\to$ store & every mechanism (10) & MP & ordered & \rfz & \rfz & \armw{Arm}-mode writer, plain: \rsn{$10^{1}$} / \rsn{$10^{1}$}; \ins{STLR};\ins{STR}: \rux / \rux; \ins{DMB LD}, \ins{DSB LD}: \rst{A2} / \rst{A2}\\
\addlinespace
load $\to$ load & every mechanism (18) & MP & ordered & \rfz & \rfz & \armw{Arm}-mode reader: \rsn{$10^{2}$--$10^{3}$} / \rsn{$10^{1}$--$10^{3}$}; \ins{DSB ST}: \rux / \rux\\
\addlinespace
load $\to$ store & every mechanism (16) & LB & ordered & \rfz & \rfz & \armw{Arm}-mode thread: \rst{A3} / \rst{A3}\\
\addlinespace
store $\to$ load & \ins{STLR};\ins{LDAR}, \ins{DMB}, \ins{DSB}, \ins{DSB ST} (4) & SB & ordered & \rfz & \rfz & plain: \rsn{$10^{5}$} / \rsn{$10^{6}$}\\
 & plain; \ins{STR};\ins{LDAPR}; \ins{STR};\ins{LDAR}; \ins{STLR};\ins{LDR}; \ins{STLR};\ins{LDAPR}; \ins{DMB ST} & SB & not ordered & \rsn{$10^{5}$--$10^{6}$} & \rsn{$10^{4}$--$10^{6}$} & \\
 & \ins{DMB LD}, \ins{DSB LD} & SB & not ordered & \rst{A2} & \rst{A2} & \\
\addlinespace
store $\to$ read-back & \ins{STR};\ins{LDR} & n7 & not ordered & \rsn{$10^{4}$} & \rsn{$10^{3}$} & \\
 & \ins{STLR};\ins{LDR} & n7 & not ordered & \rsn{$10^{4}$} & \rsn{$10^{4}$} & \\
 & \ins{STXR};\ins{LDR} & n7 & not ordered & \rsn{$10^{1}$} & \rsn{0--$10^{0}$} & \\
 & \ins{STLXR};\ins{LDR} & n7 & not ordered & \rux & \rsn{$10^{1}$} & \\
 & \ins{STR} or \ins{STLR}, then \ins{LDXR} & n7 & not ordered & \rsn{$10^{0}$--$10^{1}$} & \rsn{0--$10^{0}$} & \\
 & \ins{STLXR};\ins{LDXR} & n7 & not ordered & \rux & \rux & \\
 & \ins{STR}, \ins{STLR}, \ins{STXR} or \ins{STLXR}, then an acquire load & n7 & not ordered & \rst{A1} & \rst{A1} & \\
 & a release, then \ins{LDAR} or \ins{LDAXR} & n7 & ordered & \rfz & \rfz & \\
 & an AMO, then any load & n7 & ordered & \rfz & \rfz & \\
 & \ins{DMB LD} after the read-back & n7 & not ordered & \rst{A2} & \rst{A2} & \\
\bottomrule
\end{tabularx}
\caption{The rows of the sequential composition on the two machines, every thread in TSO mode, $6\times10^8$ iterations of each test on each machine. Entries as in \cref{tab:arm-results}. TSO orders every store $\to$ store, load $\to$ load and load $\to$ store row whatever the mechanism; their control is the same test with that thread in Arm mode, where the pair is not ordered. The store $\to$ load control is SB with plain accesses. A number in parentheses counts the mechanisms of a row. The read-back rows are n7, in which a thread stores \ins{x} and reads it back; each row gives the forms of that store and load.}
\label{tab:tsoarm-results}
\end{table}

%% file: tables/res-tso-multi.tex
\begin{table}[!htbp]
\centering
\small
\begin{tabular}{@{}lrrrll@{}}
\toprule
& \multicolumn{2}{c}{Forbidden} & \multicolumn{3}{c}{Allowed controls}\\
\cmidrule(lr){2-3}\cmidrule(lr){4-6}
Shape & tests & seen & tests & M1 & M2\\
\midrule
RWC & 4 & \rfz & 2 & \rsn{$10^{1}$--$10^{4}$} & \rsn{$10^{1}$--$10^{4}$}\\
IRIW & 4 & \rfz & 2 & \rsn{$10^{3}$} & \rsn{$10^{3}$}\\
WRC & 4 & \rfz & 1 & \rsn{$10^{4}$} & \rsn{$10^{1}$}\\
ISA2 & 8 & \rfz & 2 & \rsn{$10^{0}$--$10^{1}$} & \rsn{$10^{0}$}\\
\bottomrule
\end{tabular}
\caption{WRC, ISA2, RWC and IRIW with every thread in TSO mode, under every combination of the doubtful mechanisms (\cref{sec:method-tests}), $6\times10^8$ iterations of each test on each machine. TSO orders every edge but store $\to$ load, so a control puts one thread in Arm mode, with plain accesses; RWC also has a control with every thread in TSO mode and its store $\to$ load edge unordered. Format as in \cref{tab:arm-multi}.}
\label{tab:tso-multi}
\end{table}

%% file: tables/res-tso-rmw.tex
\begin{table}[!htbp]
\centering
\footnotesize
\begin{tabularx}{\linewidth}{@{}>{\raggedright\arraybackslash}Xlll@{}}
\toprule
Store $\to$ load pairs with an RMW, both threads in TSO mode & Sequential composition & M1 & M2\\
\midrule
an AMO as the store, as the load, or between (86) & locked & \rfz & \rfz\\
a store, then a successful LL/SC pair (6) & atomicity & \rfz & \rfz\\
\ins{STLR}, then \ins{LDAXR}; or an LL/SC pair with \ins{STLXR}, then \ins{LDAR} or \ins{LDAXR} (3) & release, acquire & \rfz & \rfz\\
an LL/SC pair, then \ins{LDR}, \ins{LDAPR} or \ins{LDAR} (5) & not ordered & \rsn{$10^{5}$--$10^{6}$} & \rsn{$10^{4}$--$10^{5}$}\\
Dekker, an AMO between store and load, every form and ordering, and a failing \ins{CAS} (13) & locked & \rfz & \rfz\\
Dekker, an LL/SC pair between store and load, each annotation (3) & not ordered & \rsn{$10^{5}$} & \rsn{$10^{4}$}\\
\bottomrule
\end{tabularx}
\caption{RMWs on TSO threads, $6\times10^8$ iterations of each test on each machine. The first four rows are every store $\to$ load pair with an exclusive or an AMO access, each in SB with \ins{DMB} on the other thread. The last two rows are hand-written, with the RMW on a third location between each thread's store and load. The controls are SB without the RMW ($10^{5}$ on the M1, $10^{6}$ on the M2) and Dekker without the RMW ($10^{6}$, $10^{5}$). For LL/SC pairs only runs in which the store-exclusives succeed are counted. Entries as in \cref{tab:arm-results}.}
\label{tab:tso-atomics}
\end{table}

%% file: sections/results-compound.tex
\section{Mixed Programs}
\label{sec:compound}

Do programs whose threads run in both modes follow the compound model? The compound model says two things: each thread keeps the preserved program order of its own mode (whatever the modes of the other threads), and the three axioms (which govern how writes propagate) are shared by all threads. We ran 529 tests that mix Arm and TSO threads. The compound model forbids the outcome of 348 of them, and the model of the machines, which adds A1--A3, forbids 377. Neither machine showed any of these 377. We first ask whether a thread's order depends on the mode of its partner, and then whether writes propagate across modes as the axioms require. Last, we take the read-back tests, the tests with RMWs, and the three orders A1--A3 that the machines add.

\subsection{Does a Thread's Order Depend on Its Partner's Mode?}
\label{sec:compound-partner}

Every row of \cref{tab:arm-results,tab:tsoarm-results} was run twice: once with an Arm-mode partner and once with a TSO-mode partner. If a thread's order does not depend on its partner's mode, the two runs should agree: the outcome appears in both or in neither. They agree in 423 of the 424 pairs of runs, 212 on each machine. The exception is two plain stores on an Arm thread on the M2: never reordered with an Arm-mode reader, and reordered 67 times in $6\times10^8$ with a TSO-mode reader. A second run of the Arm-mode test showed the outcome 26 times, so the exception is variation between runs, not a dependence on the partner's mode.

\subsection{Propagation Across Modes}
\label{sec:compound-systematic}

\input{tables/res-mixed-multi}

\Cref{tab:mixcov-results} gives the four propagation shapes, WRC, ISA2, RWC and IRIW, under every assignment of modes that mixes Arm and TSO threads (\cref{sec:method-tests}). No forbidden outcome was observed. In the shapes whose controls appear, a write of a thread in one mode thus becomes visible to all threads of the other mode at the same time. This is what the compound model mandates (\cref{sec:axioms}): the preserved program order of each thread, in its own mode, together with communication between threads, never forms a cycle.

The controls show how well each shape is exercised. RWC controls appear $10$ to $10^6$ times. WRC controls appear up to $10^5$ times on the M1 but at most tens of times on the M2. ISA2 controls are rare on both machines, at most $10^2$ times on the M1 and eight times on the M2, as when all threads run in one mode. IRIW controls appear in all but one test on the M1 and all but three on the M2. One of these, with both writers in TSO mode and both readers in Arm mode, one ordered by an address dependency and the other not, never appears on either machine; with every thread in Arm mode it appears $10^2$ to $10^4$ times. We do not know why.

\subsection{Read-Back Tests, RMWs, and A1--A3}

\paragraph{Reading a thread's own store.} \Cref{sec:tso-seq} decides which loads on a TSO thread read their own store early, with every thread in TSO mode. We also ran the same tests with Arm observers: 62 tests, n6, n7 and each store and load form. The model of the machines forbids 33 of them, and neither machine showed any. Of the 29 it allows, 27 appeared on both machines, and the other two behave as with every thread in TSO mode (\cref{tab:tsoarm-results}).

\paragraph{RMWs.} We also ran the store $\to$ load pairs of \cref{tab:tso-atomics} with the RMW on a TSO thread and an Arm partner, 100 tests. The compound model forbids 95 of the outcomes, and neither machine showed any; the five it allows, an LL/SC pair followed by a load, appeared on both machines, $10^4$ to $10^6$ times. In the 65 hand-written Dekker tests, an RMW on an Arm thread meets an RMW on a TSO thread. The compound model forbids one of their outcomes, and it was not observed.

\paragraph{A1--A3.} The three orders the machines add also hold in mixed programs. In 29 mixed tests, an order added on one thread closes a cycle with the order of a thread of the other mode. The compound model allows these outcomes and the model of the machines forbids them. None of the 29 was observed.

\paragraph{Allowed outcomes never observed.} Nine outcomes that the model of the machines allows never appeared in mixed programs. One is the IRIW case of \cref{sec:compound-systematic}. The other eight are mixed forms of outcomes that are rare, or never seen, when all threads run in one mode. We add no order for any of them.

%% file: tables/res-mixed-multi.tex
\begin{table}[!htbp]
\centering
\small
\begin{tabular}{@{}lrrrll@{}}
\toprule
& \multicolumn{2}{c}{Forbidden} & \multicolumn{3}{c}{Allowed controls}\\
\cmidrule(lr){2-3}\cmidrule(lr){4-6}
Shape & tests & seen & tests & M1 & M2\\
\midrule
RWC & 18 & \rfz & 10 & \rsn{$10^{1}$--$10^{5}$} & \rsn{$10^{1}$--$10^{6}$}\\
IRIW & 56 & \rfz & 16 & \rsn{0--$10^{3}$} & \rsn{0--$10^{4}$}\\
WRC & 27 & \rfz & 4 & \rsn{$10^{0}$--$10^{5}$} & \rsn{0--$10^{1}$}\\
ISA2 & 51 & \rfz & 8 & \rsn{0--$10^{2}$} & \rsn{0--$10^{0}$}\\
\bottomrule
\end{tabular}
\caption{WRC, ISA2, RWC and IRIW under every assignment of modes that mixes Arm and TSO, $6\times10^8$ iterations of each test on each machine. Format as in \cref{tab:arm-multi}. Where TSO orders the edge that a control leaves unordered, the control puts that thread in Arm mode, so a few controls have every thread in Arm mode.}
\label{tab:mixcov-results}
\end{table}

%% file: sections/results-scale.tex
{%
\section{Long Runs of the Added Orders}
\label{sec:scale}

Each order of \cref{sec:models-apple}, and the rule that a failing \ins{CAS} writes, rests on outcomes that never appeared in $6\times10^8$ iterations, next to controls that did (A3 has none; \cref{sec:method-reading}). Such a zero does not distinguish a forbidden outcome from a rare one; the acquire-AMO outcome of \cref{sec:arm-atomics} was such a zero and turned out to be rare. We therefore ran each of these tests $2.5\times10^{10}$ times, about 40 times as often as in the suite ($5\times10^9$ times for the three-thread tests), on three setups: the M2 and the M1 under Linux, and an M2 Max under macOS. macOS cannot put a thread in TSO mode (\cref{sec:method}), so the M2 Max runs only the tests whose threads are all in Arm mode. Threads are not pinned, and litmus7 scans memory with stride seven (\textsf{-st 7}), the setting under which the acquire-AMO outcome appeared most often.

In a two-thread Dekker, SB or LB test the outcome is possible only in iterations in which the two threads run their accesses at the same time, and in most iterations they do not. We call these iterations overlapping, and count those in which both loads read 1 (in LB, both read 0).

Two checks show that the long runs are sound. Outcomes that Arm forbids never appeared, in $5.5\times10^5$ to $9\times10^6$ overlapping iterations per test. The acquire-AMO outcome did appear, on every setup: 12 times on the M2 Max, in $3.2\times10^6$ overlapping iterations; 4 times on the M2 under Linux, in $1.5\times10^7$; and twice on the M1 under Linux, in $5.0\times10^6$. Every other zero held, in both modes and on every setup on which it ran, and every control appeared (\cref{tab:scale}).

\input{tables/res-scale}

We use the acquire-AMO outcome as a yardstick, assuming that a broken order would fail about as often per overlapping iteration. It appeared about once in $3\times10^5$ overlapping iterations on the M2 Max, and once in $3\times10^6$ under Linux. An outcome as rare would have appeared 17 to 60 times in each Arm-mode test of the added orders on the M2 Max, and about 27 and 44 times in the M1 runs of \ins{DSB LD} on TSO threads and the failing \ins{CASAL}; none appeared. The other Linux runs would have shown such an outcome only once or twice, so they add little evidence for \ins{DMB LD} on TSO threads. The yardstick does not apply to A1, whose n6 and n7 tests have no overlap count.
}

%% file: tables/res-scale.tex
\begin{table}[!htbp]
\centering
\footnotesize
\setlength{\tabcolsep}{4pt}
\begin{tabular}{@{}lllccc@{}}
\toprule
& Test & Control & M2 Max, macOS & M2, Linux & M1, Linux\\
\midrule
Arm forbids & Dekker, \ins{SWPAL} replaces the store & \ins{SWPL} & 0 / 5.2 & 0 / 8.6 & 0 / 4.9\\
 & Dekker, \ins{SWPL} replaces the load & \ins{SWPL} as store & 0 / 2.4 & 0 / 0.6 & 0 / 0.9\\
 & SB, \ins{DMB} & SB, one \ins{DMB} & 0 / 5.6 & 0 / 3.2 & 0 / 4.9\\
\midrule
Rare & Dekker, \ins{SWPA} replaces the store & \ins{SWPL} & \textbf{12} / 3.2 & \textbf{4} / 15 & \textbf{2} / 5.0\\
\midrule
A2, Arm & SB, \ins{DMB LD} & SB, one \ins{DMB} & 0 / 4.6 & 0 / 4.2 & 0 / 3.0\\
 & SB, \ins{DSB LD} & SB, one \ins{DMB} & 0 / 4.7 & 0 / 5.1 & 0 / 2.5\\
A2, TSO & SB, \ins{DMB LD} & SB, one \ins{DMB} & -- & 0 / 5.7 & 0 / 4.9\\
 & SB, \ins{DSB LD} & SB, one \ins{DMB} & -- & 0 / 6.3 & 0 / 81\\
A3 & LB, \ins{DMB} on one thread & none & 0 / 16 & 0 / 4.2 & 0 / 3.4\\
CAS & Dekker, failing \ins{CASAL} between & failing \ins{CAS} & 0 / 5.5 & 0 / 5.3 & 0 / 133\\
\midrule
A1, Arm & n7, \ins{LDAPR} & address dep. & 0 & 0 & 0\\
 & n7, \ins{LDAR} & address dep. & 0 & 0 & 0\\
 & n6, \ins{LDAPR}, Arm partner & plain load & 0 & 0 & 0\\
 & n6, \ins{LDAPR}, TSO partner & plain load & -- & 0 & 0\\
A1, TSO & n7, \ins{LDAPR} & plain load & -- & 0 & 0\\
 & n7, \ins{LDAR} & plain load & -- & 0 & 0\\
\bottomrule
\end{tabular}
\caption{The tests behind the added orders, run with unpinned threads, $2.5\times10^{10}$ times each ($5\times10^9$ for n7, which has three threads). A cell a / b gives the times the outcome appeared (a) and the overlapping iterations in millions (b); the n6 and n7 rows have no overlap count. Every control appeared on every setup on which it ran. --: not run; macOS cannot set TSO mode, so the M2 Max runs only Arm-mode tests.}
\label{tab:scale}
\end{table}

%% file: sections/conclusion.tex
\section{Conclusion}
\label{sec:conclusion}

Apple's processors let a thread switch between the Arm and TSO memory models while keeping the Arm instruction set. Specifying such a processor takes two compositions. The first is across threads: when threads run under different models, a single model must govern how they interact. The second is within a thread: a TSO thread can run Arm instructions that x86-TSO does not have, such as \ins{LDAPR}, \ins{DMB LD} and the LL/SC pair, and the model must say how each of them is ordered.

We proposed the compound model for the first and a sequential composition for the second. Under the sequential composition, every access of a TSO thread is a TSO load or store, whatever its annotation, and keeps TSO's order; AMOs are locked; and Arm's orders are added. This gives every instruction a meaning. Everything else follows from the two models; the one choice we made is to lock AMOs.

We tested the proposal on an Apple M1 and an Apple M2. Arm mode follows Arm's official model. TSO mode follows x86-TSO on the operations x86 has, and the sequential composition on the other instructions. Programs that mix the two modes follow the compound model. Neither machine showed an outcome that any of these models forbids. Both machines are also stronger than the models in two ways. They preserve three orders that the models do not require, A1--A3, and a \ins{CAS} whose comparison fails behaves as if it writes back the value it read, which makes a failing \ins{CASAL} a fence. All of these held in runs up to 40 times longer (\cref{sec:scale}; an apparent order of an acquire AMO's write before a later load did not, as the machines do reorder them, though rarely). A1--A3 and the failing \ins{CAS} describe the two machines, not what the two modes require, so we keep them out of the proposed models (\cref{sec:models-apple}).

Last, the Arm model itself orders more than its preserved program order states. A successful release RMW orders earlier stores before its read. The Arm model implies this through its atomicity axiom, although no rule of preserved program order states it.

%% file: sections/artifacts.tex
\section{Artifacts}
\label{sec:artifacts}

The artifact contains the 936 litmus tests, in litmus7 and herd form, and 18 programs in x86 assembly. A list gives, for each test, what it tests, its verdict under each model and, where it has one, its control. The artifact also contains the script that generates the tests; the herd models, \textsf{arm.cat}, \textsf{tso.cat}, \textsf{compound.cat} and their \textsf{-apple} versions; the scripts that run the tests on Apple machines; and the raw results and litmus7 output of every run.

One command checks the results. It recomputes every table and the counts quoted in the text from the raw results, recomputes every herd verdict, checks our models against herd's own Arm and x86-TSO models, and checks that each compound model reduces to its single-mode models on single-mode tests. The run scripts rerun the tests on an M1 or M2 under Asahi Linux, and the long runs also on an M2 Max under macOS. The artifact is available at \url{https://github.com/vijay-nagarajan/apple-tso-litmus}.